\documentclass[nofootinbib,aps,preprint,onecolumn,amsmath,amssymb,superscriptaddress,eqsecnum,floatfix,scrartcl]{revtex4-1}
\pdfoutput=1      
\usepackage{amssymb}  
\usepackage{color,soul} 
\usepackage{graphicx}  
\usepackage[tight]{subfigure}  
\usepackage{mathrsfs}
\usepackage{bbm}
\usepackage[pdftex,pdfusetitle,bookmarks=true,colorlinks=true,citecolor=blue,urlcolor=blue,linkcolor=magenta]{hyperref}
\usepackage{hypcap}

\begin{document}

\title{Energy trapped Ising model}    
\author{Andrea Amoretti}
\affiliation{Dipartimento di Fisica, Universita di Genova, Via Dodecaneso 33, 16146, Genova, Italy}
\affiliation{INFN, sezione di Genova, Via Dodecaneso 33, 16146, Genova, Italy}  
\email{andrea.amoretti@ge.infn.it}
\author{Gianluca Costagliola}
\affiliation{Civil Engineering Institute, Materials Science and Engineering Institute, École Polytechnique Fédérale de Lausanne (EPFL), CH-1015 Lausanne, Switzerland}
\author{Nicodemo Magnoli}
\affiliation{Dipartimento di Fisica, Universita di Genova, Via Dodecaneso 33, 16146, Genova, Italy}
\affiliation{INFN, sezione di Genova, Via Dodecaneso 33, 16146, Genova, Italy}  
\author{Marcello Scanavino}
\affiliation{Dipartimento di Fisica, Universita di Genova, Via Dodecaneso 33, 16146, Genova, Italy}
\affiliation{INFN, sezione di Genova, Via Dodecaneso 33, 16146, Genova, Italy}

\date{\today}

\begin{abstract}
In this paper we have considered the 3D Ising model perturbed with the energy operator coupled with a non uniform harmonic potential acting as a trap, showing that this  system satisfies the trap-size scaling behavior. Eventually, we have computed the correlators $\langle \sigma (z) \sigma (0)\rangle$, $ \langle \epsilon (z) \epsilon (0)\rangle$ and $\langle \sigma (z) \epsilon (0)\rangle$ near the critical point by means of conformal perturbation theory. Combining this result with Monte Carlo simulations, we have been able to estimate the OPE coefficients $C^{\sigma}_{\sigma\epsilon}$, $C^{\epsilon}_{\sigma\sigma}$ and $C^{\epsilon}_{\epsilon\epsilon}$, finding a good agreement with the values obtained in \cite{Costagliola_2016,Kos:2016ysd}.
\end{abstract}
\maketitle
\tableofcontents

\section{Introduction}
In recent years, conformal data for several CFTs have been determined thanks to the conformal bootstrap program \cite{El_Showk_2012,El-Showk:2014dwa,Gliozzi_2013,Gliozzi_2014,Kos:2016ysd}.  In addition to that, combining this numerical high-precision technique to analytical methods developed in the framework of Conformal Perturbation Theory (CPT) \cite{Guida:1995kc,Guida_1997,Amoretti:2017aze}, it is possible to determine the behavior of the off-critical correlators of many different systems. This approach has been applied successfully to the well known $3D$ Ising model, by adding perturbations proportional to the spin and the energy operator \cite{Caselle:2015csa,Caselle:2016mww}.

Starting from a slightly different perspective, CPT can be also combined to Monte Carlo simulations to get insight both on the behavior of the correlators outside the critical point and on the CFTs data at the critical point. In \cite{Costagliola_2016}, the author followed this approach to study the Ising model perturbed by a confining potential coupled to the spin operator. This model is particularly interesting because the behavior of the 1-point expectation values can be argued just by applying simple renormalization group arguments \cite{Campostrini:2009ema,Campostrini_2009,Ceccarelli_2013}, and depends only on the trap potential parameters (trap size scaling (TSS) behavior). Moreover, many experiments involving Bose-Einstein condensates and cold atoms show a critical behavior even in the presence of a trapping potential \cite{RevModPhys.74.875,RevModPhys.74.1131}.

In this paper we pursue further this program, studying the Ising model perturbed by a trapping potential coupled to the energy operator in 3 dimensions. There are many reasons to investigate this system. From a purely theoretical point of view, one can wonder if the TSS argument still holds if the trapping potential is coupled to the energy operator instead of the spin operator. Moreover, studying the effects of the energy-trapping potential on the $2$-point functions out of criticality provides an alternative method to estimate the CFT data at the critical point.

Finally, the study of the correlation functions out of the critical point is relevant also from the experimental point of view. Indeed, a trapping potential coupled to the energy operator can be effectively seen as a perturbation of the system by a non-uniform thermal gradient, a \emph{thermal trap}. This setup might be implemented in real system experiments and the knowledge of the correlators is fundamental in order to understand the behavior of the observables of this system out of the critical point.


\section{The  model and the trap size scaling}
We consider the Ising model perturbed by a confining potential coupled to the energy operator:
\begin{equation}
S=S_{cft}+\int d^d z U(z) \epsilon(z) \ ,
\end{equation}
where $S_{cft}$ is the $d$-dimensional Ising model action, $z$ is the radial coordinate and $U(r)=\rho |z|^p$ is the trap potential. In this paper we will consider $p\ge2$, focusing mostly on the harmonic potential case $p=2$. The parameter $\rho$ is the trap parameter, which is related to the characteristic trap length $l^{-p} \equiv \rho$ defined in \cite{Campostrini:2009ema}, and determines the shape of the trap. Here we will study the large-trap case, namely the small $\rho$ regime, where the CPT approach can be safely applied.

\subsection{The one point functions}
As shown in \cite{Campostrini:2009ema}, where the Trap Size Scaling (TSS) ansatz has been introduced, the behavior of the 1-point function can be extracted using renormalization group arguments. In fact, it can be shown that near the critical point the one-point functions for spin and energy, defined in the center of the trap ($z=0$), are
\begin{equation}\label{eq.1pfunction}
\langle \sigma(0)\rangle_\rho=A_\sigma \rho^{\theta\Delta_\sigma}   \ , \qquad \langle\epsilon(0)\rangle_\rho= A_\epsilon \rho^{\theta  \Delta_\epsilon} \ ,
\end{equation}
where $\Delta_\sigma$, $\Delta_\epsilon$ are the scaling dimensions of the operators $\sigma$ and $\epsilon$ respectively,  $A_\sigma$ and $A_\epsilon$ are non universal constants and the exponent $\theta$ is the characteristic trap exponent. This exponent can be determined using scaling arguments if one notices that the perturbation has to be scale invariant. Rescaling the radial coordinate $z$ by a factor $b$, $z \rightarrow \frac{z}{b}$, the perturbation transforms as 
\begin{equation}
\int d^d z' U'(z') \epsilon'(z') = b^{-d + \Delta _{\rho} -p + \Delta _{\epsilon}} \int d^d z U(z) \epsilon(z) \ ,
\end{equation}
where $ {\Delta_{\rho}} = \frac{1}{\theta}$. Eventually the scale invariance condition yields  
\begin{equation}\label{scaleinv}
\Delta _{\rho} = p + d - \Delta _{\epsilon} \ .
\end{equation}

\subsection{Two-point functions}

Regarding the two-point functions, we can make use of the Operator Product Expansion (OPE) to express them as series involving 1-point expectation values:
\begin{equation}\label{eq:OPE}
\langle O_i (z) O_j(0)\rangle_\rho=\sum_k C_{ij}^k(\rho,z)\langle O_k(0) \rangle_\rho \ .
\end{equation}
Each of the Wilson coefficient $C_{ij}^k(\rho,z)$, evaluated outside the critical point, can be expanded in series of the trap characteristic parameter $\rho$, namely:
\begin{equation}\label{eq:cpt_general}
\langle O_i (z) O_j(0)\rangle_\rho=\sum_k [C_{ij}^k(0,z)+ \rho\partial_\rho C_{ij}^k(0,z)+...]\langle O_k(0) \rangle_\rho
\end{equation}
As shown in \cite{Guida:1995kc}, the series expansion asymptotically converges and all the coefficients are infrared finite. Moreover the derivatives of the coefficients can be evaluated systematically in terms of quantities of the unperturbed theory. Before doing that, it is useful to write the fusion rules, in order to understand which of the Wilson coefficients identically vanish. Among the primary operators of the 3D Ising model, we are interested in the identity $I$ together with only two relevant ones, namely $\sigma$ and $\epsilon$. The corresponding fusion rules are:
\begin{equation}
[\sigma][\sigma]=[I]+[\epsilon]+... \ , \quad
 [\epsilon][\epsilon]=[I]+[\epsilon]+... \ , \quad [\sigma][\epsilon]=[\sigma]+... \ .
\end{equation}
These relations imply that any correlation functions containing an odd number of $\sigma$s identically vanishes. Contrary to the $d=2$ case, where Kramers-Wannier duality ($\langle[\epsilon]^n[I]^l\rangle=(-1)^n\langle[\epsilon]^n[I]^l\rangle$) implies that $C_{\epsilon\epsilon}^\epsilon=0$, in $d=3$ this Wilson coefficient is in general non-trivial and must be taken into account in the series expansions.


\subsection{Wilson coefficients in the $d=3$ case}
In three spatial dimensions, the knowledge of correlators at the critical point is limited to two and three-point functions, and the scaling dimensions and structure constants have been evaluated numerically in \cite{Kos:2016ysd}: $(\Delta_\sigma,\Delta_\epsilon)= (0.5181489(10), 1.412625(10))$ and $ (C^{\epsilon}_{\sigma \sigma},C^{\epsilon}_{\epsilon\epsilon})=(1.0518537(41), 1.532435(19))$. Out of the critical point, the correlators can be expanded as a series of the parameter $\rho$ in the following way:
\begin{eqnarray}
\langle \sigma(z_1) \sigma(0) \rangle_{\rho}&=& C^{I}_{\sigma \sigma} (z_1)+C^{\mathcal{\epsilon}}_{\sigma \sigma} (z_1) A_{\epsilon} \rho^{\theta\Delta_\epsilon}+ \rho \partial_{\rho} C^{I}_{\sigma \sigma}(z_1)+... \ ,\label{eq:3d_ss}\\
\langle \epsilon(z_1) \epsilon(0) \rangle_{\rho}&=& C^{I}_{\epsilon\epsilon}(z_1)+C^{\epsilon}_{\epsilon\epsilon}(z_1)  A_{\epsilon} \rho^{\theta\Delta_\epsilon}+ \rho \partial_{\rho} C^{I}_{\epsilon\epsilon} (z_1) +... \  ,\label{eq:3d_ee}\\
\langle \sigma(z_1) \epsilon(0) \rangle_{\rho}&=& A_{\sigma}\rho^{\theta \Delta_{\sigma}}( C^{\sigma}_{\sigma \epsilon} (z_1)+ \rho \partial_{\rho} C^{\sigma}_{\sigma \epsilon}(z_1)+...) \ .\label{eq:3d_se}
\end{eqnarray}
As said before, the derivatives of Wilson coefficient can be written in terms of known quantities \cite{Guida:1995kc,Guida_1997,Amoretti:2017aze}. For instance, $\partial_{\rho}C^{I}_{\sigma\sigma}(z_1)$ reads:
\begin{equation}
- \partial_{\rho} C^{I}_{\sigma \sigma} (z_1)=
 \int d^3z_2 \ |z_2|^p \Big[\langle \sigma(z_1) \sigma(0) \epsilon(z_2) \rangle-C^{\epsilon}_{\sigma \sigma}(z_1) \langle\epsilon(z_2) \epsilon(0) \rangle \Big] \ .\label{eq:deCss1_1}
\end{equation}
This integral can be evaluated using a Mellin transform technique (see appendix \ref{appendix} for details).
In particular, the second term is just a regulator needed to cancel the IR-divergent part, meaning that only the first term contributes to
the final result. Expanding the first term in \eqref{eq:deCss1_1} in terms of the known correlation function at the critical point we find:
\begin{equation}
\partial_{\rho}C^{I}_{\sigma\sigma}(z_1)=-z_1^{\Delta_t-2\Delta_\sigma+p} C^\epsilon_{\sigma\sigma}\int d^3y \frac{ y^p}{y^{\Delta_\epsilon}(1+y^2-2y\cos\theta)^{\frac{\Delta_\epsilon}{2}}} \ , \label{eq:deCss1_2}
\end{equation}
where $\Delta_t=3-\Delta_\epsilon$ and $y=z_2/z_1$. We refer to Appendix \ref{appendix} for the details of the computation. The final result is:
\begin{equation}
\partial_{\rho}C^{I}_{\sigma\sigma}(z_1)=-z_1^{\Delta_t-2\Delta_\sigma+p} C^\epsilon_{\sigma\sigma} I(p)  \ ,
\end{equation}
where  $I(p)$ is numerical factor that can be expressed in terms of Gamma functions and the relevant parameters of the model, as shown in \eqref{Ipcomplete}. In what follow we are going to consider mostly the harmonic potential case, $p=2$, for which  $I(2)\simeq -8.4448$. 

Following the same procedure we can also evaluate the derivative of $C_{\epsilon\epsilon}^I$:
\begin{equation}
\partial_{\rho}C^{I}_{\epsilon\epsilon}(z_1)=-z_1^{\Delta_t-2\Delta_\epsilon+p} C^\epsilon_{\epsilon\epsilon} I(p) \ .
\end{equation}
Putting all together, the expressions \eqref{eq:3d_ss}-\eqref{eq:3d_se} can be expressed as:
\begin{eqnarray}
z_1^{2\Delta_\sigma}\langle \sigma(z_1) \sigma(0) \rangle_{\rho} &=& 1+C^{\epsilon}_{\sigma \sigma} A_{\epsilon} (\rho^{\theta} z_1)^{\Delta_\epsilon}- C^{\epsilon}_{\sigma \sigma}\rho z_1^{\Delta_t+2}I(2)+... \ ,  \\
z_1^{2\Delta_\epsilon}\langle \epsilon(z_1)\epsilon(0)\rangle_{\rho} &=& 1+C^{\epsilon}_{\epsilon\epsilon} A_{\epsilon} (\rho^{\theta} z_1)^{\Delta_\epsilon}- C^{\epsilon}_{\epsilon\epsilon}\rho z_1^{\Delta_t+2}I(2)+... \ , \\
z_1^{\Delta_\epsilon}\langle \sigma(z_1)\epsilon(0)\rangle_{\rho} &=&A_{\sigma} \rho^{\theta\Delta_\sigma} \bigl( C^{\sigma}_{\sigma\epsilon} +\# \rho z_1^{\Delta_t+2}+... \bigr) \ .
\end{eqnarray}

As usual in this approach, the  asymptotic convergence of the series expansion is guaranteed for distances (measured from the center of the trap) less than about one correlation length. In the last equation the symbol $\#$ stands for the numerical value of the coefficient $\partial_\rho C^{\sigma}_{\sigma \epsilon}(z_1)$. The computation of this coefficient within the CPT framework involves the use of a 4-point correlation function at the critical point:
\begin{multline}
\label{4point}
-\partial_{\rho} C^{\sigma}_{\sigma\epsilon}(z_1) \lim_{|z_3| \rightarrow \infty} \langle \sigma(z_3) \sigma(0) \rangle=\\
\lim_{|z_3| \rightarrow \infty} \int_{|z_2|<|z_3|} d^3z_2 \ |z_2|^p \Big[\langle \sigma(z_1) \sigma(z_3)\epsilon(z_2) \epsilon(0) \rangle-C^{\sigma}_{\sigma \epsilon}(z_1) \langle \sigma(z_3) \sigma(0) \epsilon(z_2)\rangle \Big]  \ .
\end{multline}
Since $\langle \sigma(z_1) \sigma(z_3)\epsilon(z_2) \epsilon(0) \rangle$ is not known analytically at the critical point, \eqref{4point} can not be evaluated exactly. However, as we will show later, this term can not be neglected and it will be determined a posteriori using Monte Carlo simulations. 



\section{Conversion to the lattice and numerical results}
The model previously described can be solved on a lattice in order to verify the validity of the CPT expansion and to get some insights in the numerical factors which we have not been able to determine analytically. The Hamiltonian of the system on a  cubic lattice can be expressed in the following form:
\begin{equation}\label{eq:trap_ising}
\mathcal{H}=-J \sum_{\langle i j \rangle}\sigma_i \sigma_j(1+U(r_i)) +h \sum_i \sigma_i 
\end{equation}
where $\sigma_{i}$ is the spin field, $r_{i}$ is its distance from the center of the confining potential and $h$ is a possible magnetic field perturbation whose importance will be shortly explained. The conformal point is recovered for $h=0$.
To get a more precise physical intuition about the trapping effect, it is convenient to perform the transformation $\sigma_i=1-2\rho_i$. Then, the new variable $\rho_i$ can only assume two values (0 and 1) and it can be thought as a density of particles in a $d$-dimensional gas. Eventually, the Hamiltonian reads:
\begin{equation}\label{trnasformedham}
\mathcal{H}=-4J \sum_{\langle i j \rangle}\rho_i \rho_j -\mu \sum_i \rho_i + 4J\sum_{\langle i j \rangle}U(r_{i})\rho_i(1-\rho_j)
\end{equation}
where $\mu=2h-4qJ$ is the chemical potential and $q$ is the coordination number ($q=6$ in three dimensions). The main advantage of this transformation is that, since the potential $U(r_{i})$ diverges at large $r_i$, it makes it apparent that the only way to prevent the last term in \eqref{trnasformedham} to diverge is to set either $\langle\rho_i\rangle=1$ or $\langle\rho_i\rangle=0$ for all $i$ far from the center of the trap. The first condition is not physically acceptable (all the particles running away to infinity) and it can be avoided by inserting a small and positive magnetic field $h$ in eq. \ref{eq:trap_ising}, namely:
\begin{equation}
\lim_{h\rightarrow0^+}\lim_{|r|\rightarrow\infty}\langle \sigma_r\rangle=1 \ .
\end{equation}
This leaves us only with the second possibility, which is equivalent to require a null density of particles ($\langle\rho_i\rangle=0$) far from the center of the lattice, which means that the system is trapped.

\subsection{Lattice implementation}

The Monte Carlo simulation is performed with the Metropolis algorithm on a cube with side $L$ and fixed boundary. The trap is centered in the middle point of the cube. The spin $i$ located on the lattice at distance $r$ from the center is denoted with $\sigma_{r_{i}}^{lat}$. 
We calculate the following observables: the spin one-point function on the central site $\langle
\sigma_0^{lat} \rangle$,  and the energy one-point function in the middle of the lattice, defined as $\langle\epsilon_{0}^{lat} \rangle
\equiv \langle \sigma_{0}^{lat}\sigma_{1}^{lat} \rangle - E_{cr}$, where $E_{cr}$ is the energy bulk contribution at the critical
point and $\langle...\rangle$ is the statistical average.

The correlation functions are calculated from the central site of the lattice up to the distance $r$ on the
central axis, averaging between the six orthogonal directions. Thus, they are defined as:
\begin{eqnarray}
G_{\sigma \sigma} (r) &\equiv & \frac{1}{6} \langle \sum_{i=1}^{3} \sigma_{0}^{lat} \; (\sigma_{r_i}^{lat} + \sigma_{-r_i}^{lat}
) \rangle  \label{eq_corr1} \ ,\\
G_{\epsilon \epsilon} (r) &\equiv & \frac{1}{6} \langle \sum_{i=1}^{3} \epsilon_{0}^{lat} \; (\epsilon_{r_i}^{lat} +\epsilon_{-r_i}^{lat} ) \rangle \label{eq_corr2} \ , \\
G_{\sigma \epsilon } (r) &\equiv & \frac{1}{6} \langle \sum_{i=1}^{3} \epsilon_{0}^{lat} \; (\sigma_{r_i}^{lat} +
\sigma_{-r_i}^{lat} ) \rangle \label{eq_corr3} \ , 
\end{eqnarray}

As the system breaks translational invariance, we may wonder $G_{\sigma \epsilon }$ to be different from $G_{\epsilon \sigma }$. However, we have verified that the differences between the two correlators are negligible within the parameter range used in the simulations and we will eventually focus on $G_{\sigma \epsilon }$ in the rest of our analysis.

We have performed our simulations focusing on the harmonic trap case, namely setting $p=2$. Moreover we have fixed the following constants to their known Ising model values: the energy bulk value $E_{cr}=0.3302022
(5)$ and the critical temperature $\beta_{c} = 0.22165462 (2) $ \cite{Hasenbusch:2012spc}, the scaling dimensions
$\Delta_\sigma = 0.51815 (2) $ and $\Delta_\epsilon = 1.41267 (13)$ \cite{El-Showk:2014dwa}. Thus, $p\theta =
2/(5-\Delta_\epsilon) \simeq 0.55752$. The uncertainty on these constants is negligible with respect to our numerical precision.

The simulations have been performed with a lattice side $L=480$ that is large enough to avoid finite size effects within our current precision. Since our observables are closely sampled around the center of the trap, we adopt a hierarchical upgrading scheme \cite{Caselle_2003}: instead of performing the Monte Carlo sweep on the whole lattice at each step, sweeps are performed in nested cycles over smaller cubic boxes of increasing size centered in the middle of the lattice. With this procedure computational times are reduced without affecting local central observables. In a single Monte Carlo simulation, starting from a configuration with all spins aligned, $5\cdot10^{6}$ sweeps have been performed, with about $10^{4}$ sweeps for thermalization. Observable uncertainties have been calculated by using the batched mean method. Moreover, final results of all observables have been obtained by averaging about 100 repeated and independent Monte Carlo simulations.

\subsection{One-point functions}
\label{sec:onepoint}
Since the potential is coupled to the temperature, which in the lattice is non-zero at the critical point, the effective scaling parameter on the lattice to be compared with analytical prediction is $\rho_{lat} \equiv \beta_{c} \rho$. Thus, the one-point functions on the lattice are:
\begin{eqnarray}
\langle \sigma_{0}^{lat} \rangle  &=&  A_{\sigma}^{lat} \; \rho_{lat}^{\theta \Delta_\sigma} \ , \label{mtrap_l}\\
\langle \epsilon_{0}^{lat} \rangle   &=&   A_{\epsilon}^{lat} \; \rho_{lat}^{ \theta \Delta_\epsilon } \ . \label{etrap_l}
\end{eqnarray}

\begin{figure}[!h]
\begin{center}
\includegraphics[scale=0.45]{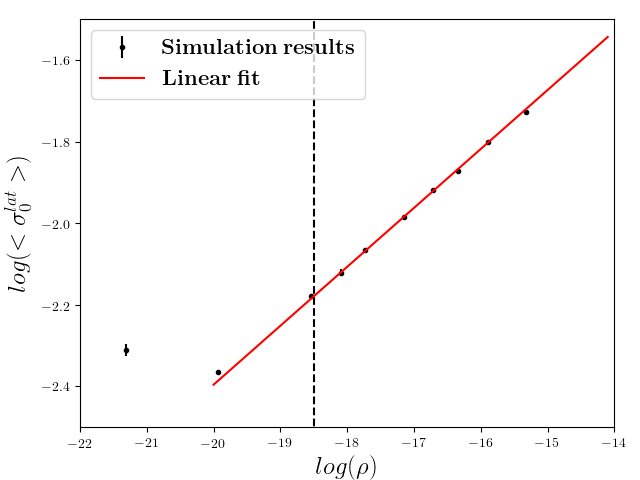}
\includegraphics[scale=0.45]{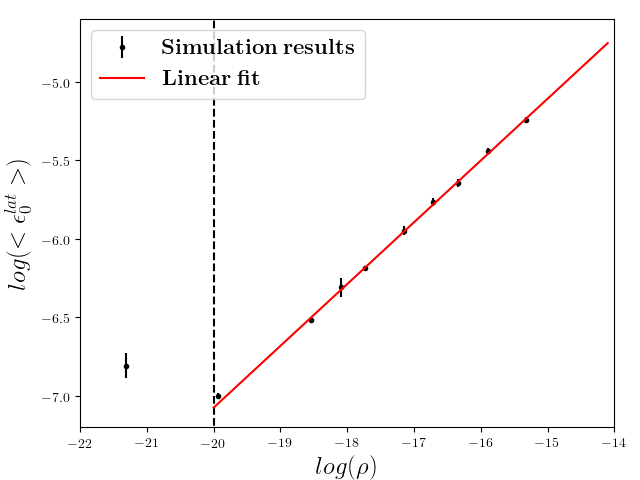}
\caption{Bi-log plots of the the spin (Left panel) and energy (Right panel) one-point functions against power law fits (red line). Due to numerical accuracy, the fits have been performed for values of $\rho$ greater than the ones indicated by the dashed vertical lines.}
\label{fig1}
\end{center}
\end{figure}
\begin{table}[h!]
\begin{center}
\begin{tabular}{|c|c|c|c|c|}
\hline
exponent & theory & simulation & $\chi^{2}/d.o.f. $ \\
\hline
$ \theta \Delta_\sigma$  & 0.144439(5) & 0.144(1)   & 0.6 \\
\hline
$ \theta \Delta_\epsilon$  & 0.39379(5) & 0.392(4)  & 1.4 \\
\hline
\end{tabular}
\caption{Exponents extracted from the fits shown in figure \ref{fig1}.}
\label{tab1}
\end{center}
\end{table}

The results for the spin and energy one-point functions are shown in figure \ref{fig1}. The fit has been performed in the range
$ 10^{-8} \leq \rho \leq 5.625\times 10^{-7} $. Within this range, the scaling exponents are in good agreement with the theoretical result predicted by the TSS argument \ref{mtrap_l}-\ref{etrap_l}, as shown in table \ref{tab1}. This confirms the validity of the TSS ansatz \cite{Campostrini:2009ema} also in the present case. Eventually, we fix the exponents to the value \ref{mtrap_l}-\ref{etrap_l} and we repeat the fit with only two free parameters to find the remaining constants, obtaining $A_{\sigma}^{lat} = 1.6390(13) $ and $A_{\epsilon}^{lat}=2.226(11)$.\\

\subsection{Two-point functions}
With the definitions \ref{eq_corr1}-\ref{eq_corr3} at hand, and taking into account the bulk contribution to the energy operator on the lattice $E_{cr}$, the two-point functions on the lattice (denoted with with the average $\langle... \rangle_{lat}$) assume the following form:
\begin{eqnarray}
\langle \sigma(r) \sigma(0) \rangle_{lat} & =& G_{\sigma \sigma} (r)  \ , \label{eq:obs1} \\
\langle \epsilon(r)\epsilon(0) \rangle_{lat} & = & G_{\epsilon \epsilon} (r) + E_{cr}^2 - E_{cr}(\langle\epsilon_{r}^{lat}\rangle + \langle\epsilon_{0}^{lat}\rangle) \ , \label{eq:obs2} \\
\langle\sigma(r) \epsilon(0) \rangle_{lat}  &=& G_{\sigma \epsilon} (r) - E_{cr} \langle\sigma_{r}^{lat}\rangle \ . \label{eq:obs3}
\end{eqnarray}

In order to make contact with the CPT theoretical results \eqref{eq:3d_ss}-\eqref{eq:3d_se}, we must consider the lattice conversion factors $R_{\sigma}$ and $R_{\epsilon}$. Regarding the first,   $R_{\sigma} = 0.55245(13)$ according to \cite{Herdeiro_2017}. Estimates of $R_{\epsilon}$ vary from 0.2306(38) \cite{Herdeiro_2017} to 0.2377(9) \cite{Costagliola_2016}. This is the largest source of systematic uncertainty in our simulations. For this reason we will adopt the average $R_{\epsilon}=0.2341$ with a variation $\pm 0.0030$ to evaluate the final systematic error. Finally, the structure constant on the lattice $(C_{\sigma \sigma}^{\epsilon})^{lat}$ must be converted taking into account the conversion rules for $\epsilon$ and $\rho$, namely $\langle\epsilon^{lat}\rangle = R_{\epsilon}\langle\epsilon\rangle $ and $\rho_{lat} = R_{\epsilon}^{-1} \rho$.
Eventually, combining \eqref{eq:3d_ss}-\eqref{eq:3d_se} with \eqref{eq:obs1}-\eqref{eq:obs3} we obtain:
\begin{eqnarray}
\langle \sigma(r) \sigma(0) \rangle_{lat}  &=& \frac{R_{\sigma}^2}{r^{2 \Delta_{\sigma}}} \left( 1 + C_{\sigma \sigma}^{\epsilon} \: R_{\epsilon}^{-1} A_{\epsilon}^{lat} \rho_{lat}^{\theta \Delta_\epsilon} r^{\Delta_\epsilon} - C_{\sigma \sigma}^{\epsilon} I(2) R_{\epsilon} \rho_{lat} r^{2+\Delta_t}   \right) \ , \label{eq_oo2} \\
\langle \epsilon(r)\epsilon(0) \rangle_{lat} &=& \frac{R_{\epsilon}^2}{r^{2 \Delta_{\epsilon}}} \left( 1 + C_{\epsilon \epsilon}^{\epsilon} \: R_{\epsilon}^{-1} A_{\epsilon}^{lat} \rho_{lat}^{\theta \Delta_\epsilon} r^{\Delta_\epsilon} - C_{\epsilon \epsilon}^{\epsilon} I(2) R_{\epsilon} \rho_{lat} r^{2+\Delta_t}   \right) \ , \label{eq_ee2}\\
\langle\sigma(r) \epsilon(0) \rangle_{lat} &=& \frac{R_{\epsilon}R_\sigma \rho_{lat}^{\theta \Delta_\sigma}}{r^{ \Delta_{\epsilon}}} \left( C_{\sigma \epsilon}^{\sigma} \: R_{\sigma}^{-1} A_{\sigma}^{lat} +b  \rho_{lat} r^{2+\Delta_t}   \right) \ . \label{eq_oe2}
\end{eqnarray}
\begin{table}[h]
\begin{center}
\begin{tabular}{|c|c|c|c|c|}
\hline
$\rho$ & range $r$ & $C_{\sigma \sigma}^{\epsilon}$ & $\chi^{2}/d.o.f. $ \\
\hline
$1\times10^{-8}$   & 7-13 & 1.059(20)[40] & 3.5 \\
\hline
$4\times10^{-8}$ & 7-13 & 1.049(5)[15]  & 0.9  \\
\hline
$9\times10^{-8}$ & 7-13 & 1.043(3)[14] & 0.2 \\
\hline 
\end{tabular}
$\qquad$
\begin{tabular}{|c|c|c|c|c|}
\hline
$\rho$ & range $r$ & $C_{\epsilon \epsilon}^{\epsilon}$ & $\chi^{2}/d.o.f. $ \\
\hline
$1\times10^{-8}$   & 6-13 & 1.46(15)[30] & 0.9 \\
\hline
$4\times10^{-8}$  & 6-13 & 1.58(7)[24]  & 0.6  \\
\hline
$9\times10^{-8}$  & 6-13 & 1.50(8)[20] & 1.1 \\
\hline 
\end{tabular}
\caption{Results of the structure constant found by fitting the data with the function \ref{eq_oo2} for various trap sizes $\rho$. The number in round brackets denotes the statistical uncertainty of the fit, while the number in square brackets denotes the systematic error due to the uncertainty of the constants. Regarding the correlator related to the table on the right side, we have sampled all the distances in the same simulation, so that statistical errors have been estimated by means of the jack-knife technique.}
\label{tab2}
\end{center}
\end{table}
The parameter $b$ in the second term of \eqref{eq_oe2} is related to the coefficient \eqref{4point}, which, as already mentioned, we have not been able to compute analytical using CPT. This parameter will be evaluated a posteriori by fitting the numerical results. 

\begin{figure}[h]
\begin{center}
\includegraphics[width=8.1cm,keepaspectratio]{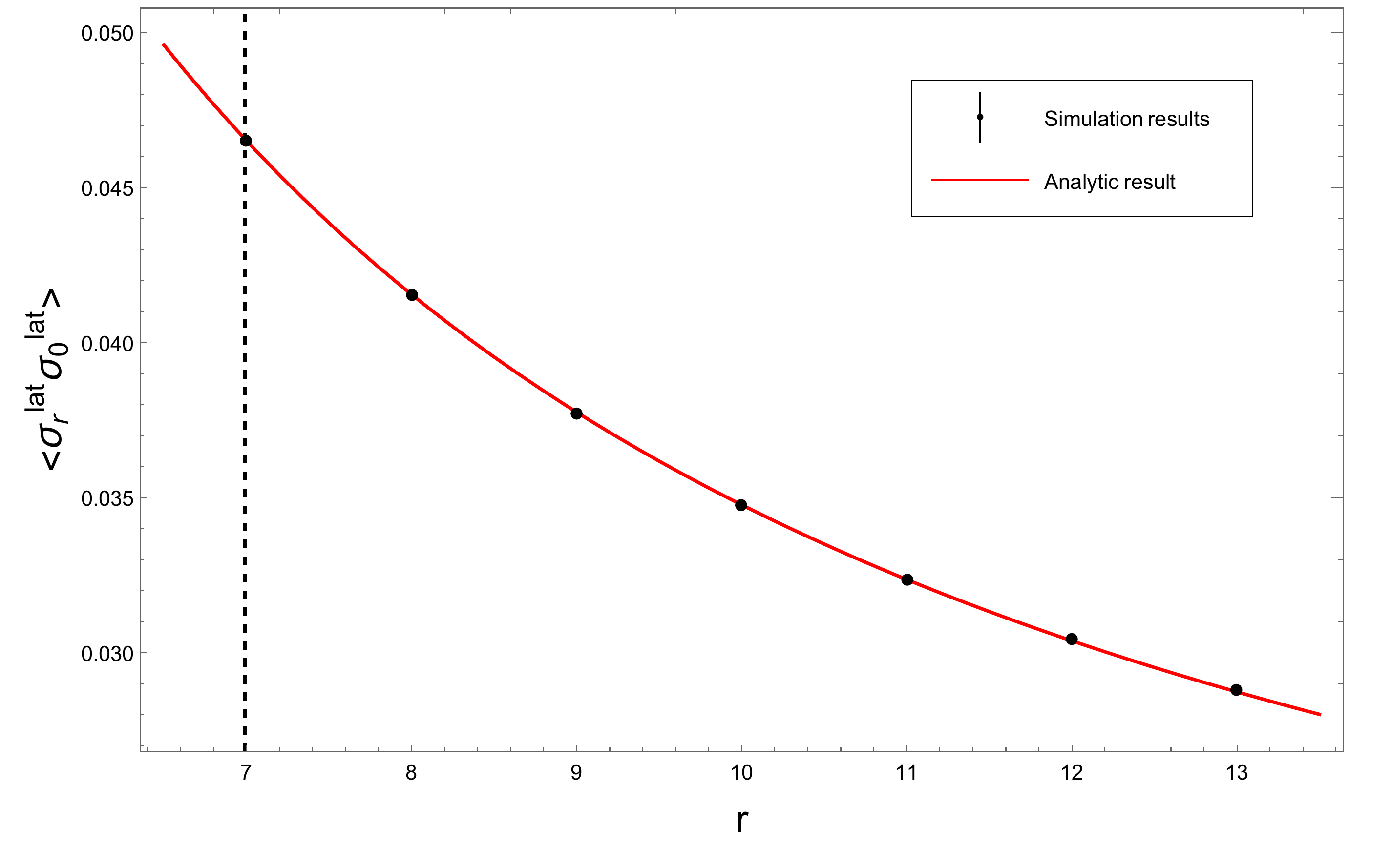}
\includegraphics[width=8.1cm,keepaspectratio]{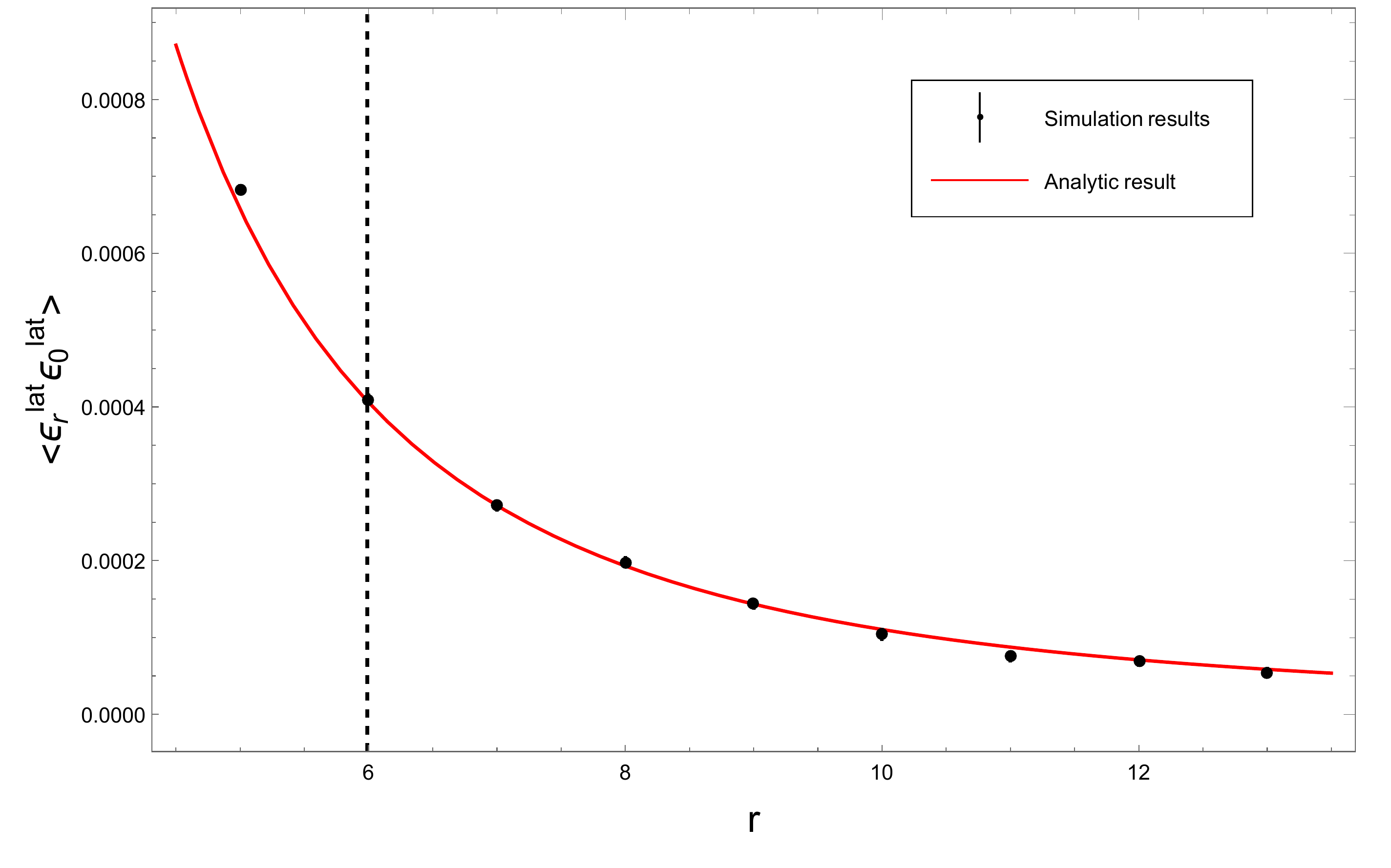}
\caption{Results for the spin-spin (left) and energy-energy (right) two-point functions and their fits with the expected behavior (eq. \ref{eq_oo2} and \ref{eq_ee2}) red line. Due to numerical accuracy, fits are performed for values of $r$ greater than the ones indicated by the black dashed vertical lines. Error bars are small and usually hidden within the points size.}
\label{fig2}
\end{center}
\end{figure}

We can now insert the lattice quantities $A_{\epsilon}^{lat}$ and $A_{\sigma}^{lat}$ calculated in section \ref{sec:onepoint}, and directly fit the continuum structure constants $C_{\sigma \sigma}^{\epsilon}$ and $C_{\epsilon \epsilon}^{\epsilon}$. Fit results, reported in the table \ref{tab2}, are in good agreement with the known values: $C_{\sigma \sigma}^{\epsilon} = 1.0518537$, $C_{\epsilon \epsilon}^{\epsilon} = 1.532435$ \cite{El-Showk:2014dwa}.
Figure \ref{fig2} shows the behavior of the correlators. More specifically, data and fits are outlined for $\rho=9\times10^{-8}$, and Monte Carlo data reproduce well the expected behavior. We obtained very similar results for the other trap-sizes reported in the tables \ref{tab2}.

\begin{figure}[h]
\begin{center}
\includegraphics[scale=0.33]{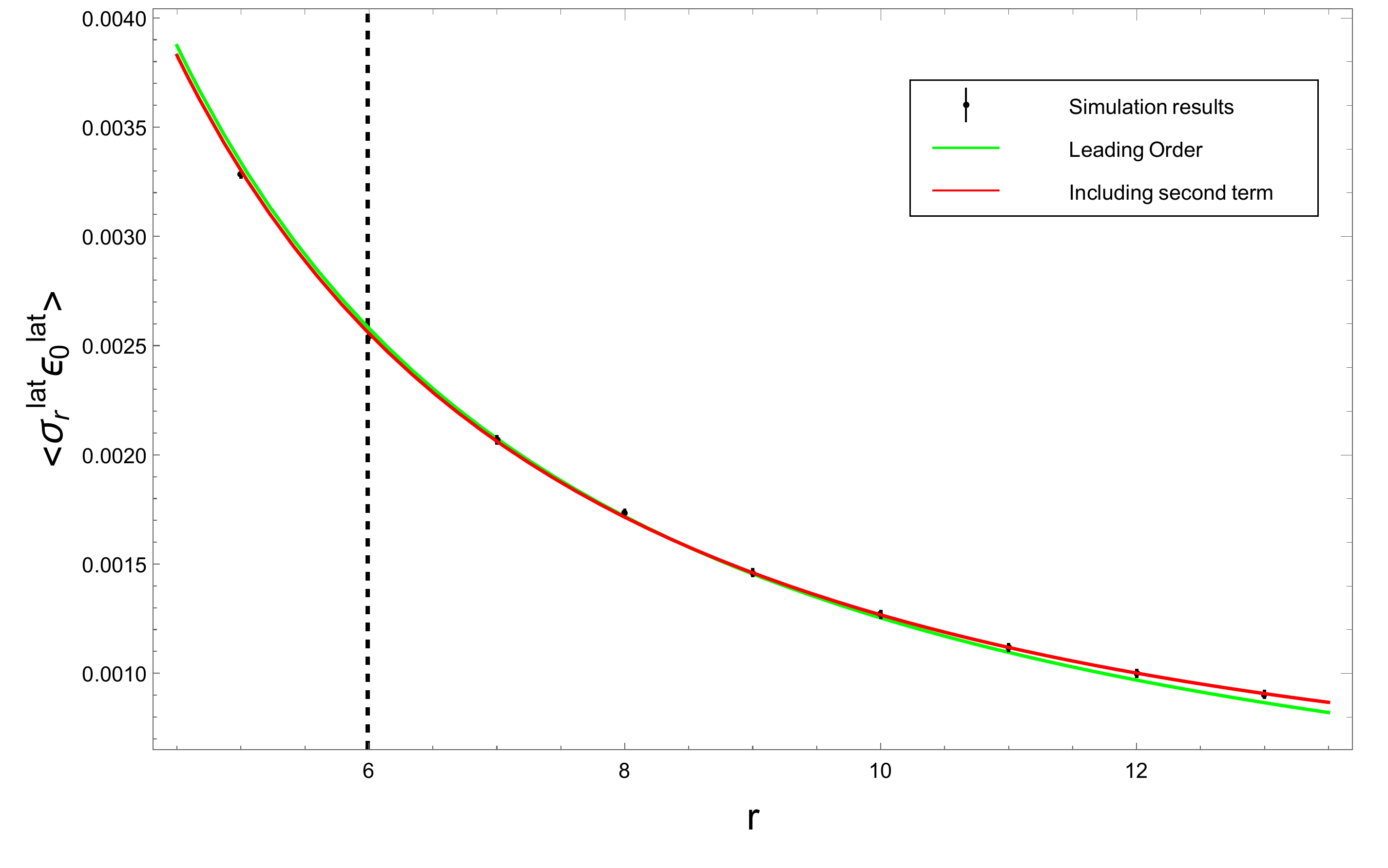}
\caption{Fit for the spin-energy two-point function against the theoretically expected behavior performed not including (green line) and including (red line) the second term in Eq. \ref{eq_oe2}. As it is evident, the red line agrees consistently better with the numerical result.}
\label{fig3}
\end{center}
\end{figure}

Table \ref{tab4} shows the fit results for the mixed correlator $\langle\sigma\epsilon\rangle$ without including the second term in \eqref{eq_oe2} ($b=0$), while table \ref{tab5} shows the fits including $b$ as a free parameter. As one can see from the tables, once $C_{\sigma\epsilon}^\sigma$ is left as a free parameter its value agrees better with the known result if we take into account the parameter $b$. This is confirmed in Figure \ref{fig3}, where it is evident that the presence of $b$ significantly improves the agreement between the theoretical prediction and the numerics. This proves that the second term in \eqref{eq_oe2} is definitely important and must be taken into account.

\begin{table}[h!]
\begin{center}
\begin{tabular}{|c|c||c|c||c|c||}
\hline
$\rho$ & range $r$ & $b$ ($C_{\sigma \epsilon}^{\sigma} = 1.0518537$) & $\chi^{2}/d.o.f. $ & $C_{\sigma \epsilon}^{\sigma}$ ($b$=0) & $\chi^{2}/d.o.f. $ \\
\hline
$1\times10^{-8}$  & 7-13 & 1.1(2)[3]$\cdot 10^{4}$ & 2.2 & 1.098(3)[10] & 0.3    \\
\hline
$4\times10^{-8}$  & 6-13 & 2.8(4)[5]$\cdot 10^{3}$  & 0.98 & 1.082(6)[12] & 1.6  \\
\hline
$9\times10^{-8}$  & 6-13 & 1.9(2)[4]$\cdot 10^{3}$ & 5.5 &  1.094(10)[14] & 4.3  \\
\hline 
\end{tabular}
\caption{Fit performed including the second term of Eq. \ref{eq_oe2} and fixing $C_{\sigma\epsilon}^\sigma$ to the known value (third column), and fit of the structure constant $C_{\sigma\epsilon}^\sigma$ setting $b=0$ (fifth column). It is evident from the data that $b$ contributes non-trivially to the correlator, as our numerical results do not match the known value for $C_{\sigma\epsilon}^\sigma$ if we set $b=0$. }
\label{tab4}
\end{center}
\end{table}

\begin{table}[h!]
\begin{center}
\begin{tabular}{|c|c|c|c|c|}
\hline
$\rho$ & range $r$ & $C_{\sigma \epsilon}^{\sigma}$ & b & $\chi^{2}/d.o.f. $\\
\hline
$1\times10^{-8}$   & 7-13  & 1.098(4)[10] & $\sim 0$ & 0.3  \\
\hline
$4\times10^{-8}$  & 6-13  & 1.067(7)[12] & 1.8(5)[1]$\cdot10^{3}$ & 0.3 \\
\hline
$9\times10^{-8}$  & 6-13 & 1.080(9)[12] & 1.0(2)[1]$\cdot10^{3}$ & 0.8 \\
\hline 
\end{tabular}
\caption{Performing the fit with two free parameters the situation improves and the numerical values of $C_{\sigma \epsilon}^\sigma$ are in agreement with the expected one within the numerical error. The results for $\rho=1\times10^{-8}$ are probably affected by some finite-size effect, as it can been seen in the $\langle\sigma\sigma\rangle$ correlator as well.}
\label{tab5}
\end{center}
\end{table}

\section{Discussion}
In this paper we have further developed the program of studying systems in their off-critical scaling regime, using the consolidated approach based on the OPE and the possibility of expanding the Wilson coefficients in terms of the perturbing parameter by means of conformal perturbation theory \cite{Guida:1995kc,Guida_1997,Amoretti:2017aze}. This has been done for the 3D Ising model perturbed by a trapping potential coupled to the energy operator. Nevertheless, the procedure can be applied in principle to other systems in a different universality class since the method only requires scale invariance at the critical point.

We have evaluated the first leading terms in the expansions of the correlators comparing the analytic predictions against numerical Monte Carlo simulations. The results for the 1-point functions outlined in Fig. \ref{fig1} confirm once again the validity of the TSS ansatz \cite{Campostrini:2009ema} as a powerful tool to determine the behavior of the expectation values of the model outside the critical point.

Despite the necessity of using large size traps and consequently large lattices, the estimates of the structure constants shown in tables \ref{tab2}  are in good agreement with the known results found in literature. This fact shows the reliability of the approach and confirms that this method is a promising tool for studying different systems out of criticality. Additionally, we have proven that the behavior of the $\langle \sigma \epsilon \rangle$ correlator is influenced by the presence of a term which depends, according to the CPT approach, on an integral involving a 4-point function \eqref{4point}. Due to the lack of knowledge on the 4-point function at the critical point in the 3D Ising model, this integral can not be evaluated analytically. However, using the high quality CFT data in \cite{Simmons-Duffin:2016wlq} one could compute the 4-point function needed in \eqref{4point} analogously to what has been done for the $\sigma$ 4-point function in \cite{Rychkov:2016mrc}. The integral \eqref{4point} may be eventually evaluated applying the procedure used to compute similar integrals involving the $\epsilon$ 4-point function in \cite{Komargodski:2016auf} and the $\sigma$ 4-point function in \cite{Behan:2017emf}. This will allow us to validate our numerical estimation (Table \ref{tab4} and \ref{tab5}) and constitutes a worthwhile future direction.

Finally, the method used in this paper could be applied to other interesting examples like systems with a quantum critical point, systems where the critical point is broken by a lattice operator (to compare with what has been done in \cite{Amoretti:2019cef} by means of AdS/CFT techniques) and the $3D$ $O(N)$ model. In particular, the latter needs to be treated carefully, because it exhibits spontaneous symmetry breaking and the dynamics of the Goldstone bosons might have a non-trivial effect on the system.


\section*{Acknowledgments} 
The project has been partially supported by the INFN Scientific Initiative SFT: “Statistical Field Theory, Low-Dimensional Systems, Integrable Models and Applications”.
 
\appendix\
\section{Mellin transform technique}\label{appendix}
The integral \ref{eq:deCss1_1} can be evaluated using a Mellin transform technique, following what has been done in \cite{Guida:1995kc,Guida_1997,Amoretti:2017aze}. In particular, it is convenient to introduce the quantity
\begin{equation}
I(m) = \int d^3 z \Theta (m |z| ) g(z) \ ,
\end{equation}
where $\Theta(|m|z)=e^{-m|z|}$ is an IR-regulator needed to guarantee the convergence of the integral. We are interested in the $m\sim 0$ expansion of $I(m)$, that can be recovered by considering its Mellin transform. Assuming that the leading behavior of $I(m)$ as  $m \rightarrow 0$ is $m^{a}$, while it approaches $m^{-b}$ when $m \rightarrow \infty$, the Mellin transform $\tilde I (s)$ is defined on the strip $-a < Re(s) < b$ in the complex $s$ plane as:
\begin{equation}
\tilde{I} ( s ) = \int_0^\infty \frac{dm}{m} m^s I( m ) \ . 
\end{equation}
Eventually, it can be proven that only the first term of the integral \ref{eq:deCss1_1} contributes, while the second one leads to a null strip so that the transform is not well defined.

The asymptotic expansion of the original function at $m=0$ is in a one to one correspondence with the poles of the Mellin transform, namely:
\begin{equation}\label{eq:mellin}
I(m) = \sum _i Res( \tilde{I} (s) )_{s = -a_i} m^{-s} \ ,
\end{equation}
where $a_1 \equiv a < a_2<...$ are the powers of $m$ in the asymptotic expansion of $I(m)$ at $m \sim 0$. \eqref{eq:mellin} tells us that we can get the corrections to the Wilson coefficients by taking the residue of the perturbative expansions at $s = 0$ if the infrared counter-terms do not give any finite contribution.

With our choice of the regulator, the Mellin transform of $I(m)$ can be easily obtained by using the convolution theorem, finding:
\begin{equation}
\tilde{I} (s) = \Gamma (s) \tilde{g} (1-s) \ ,
\end{equation}
where
\begin{equation}
\tilde {g} (1-s) = \int d^3 z |z|^{-s} g(z) \ ,
\end{equation}
is essentially the Mellin transform of $g$ up to angular coefficients. This means that in order to find an expression for the derivatives of the Wilson coefficients, one just needs to evaluate the Mellin transform of the function $g(z)$.

In our case, as it can be seen from Eq. \ref{eq:deCss1_2},
\begin{equation}
g(z)= \frac{ z^{p-\Delta_\epsilon}}{(1+z^2-2z\cos\theta)^{\frac{\Delta_\epsilon}{2}}} \ .
\end{equation}
The Mellin transform can be evaluated by performing the angular integral and rewriting the result in terms of beta-functions as follows:
\begin{multline}
\tilde{I}(s)=\Gamma(s)\frac{2\pi}{2-\Delta_\epsilon}\Bigl[ B(p+2-\Delta_\epsilon-s,2\Delta_\epsilon-4-p+s) +\\-B(p+2-\Delta_\epsilon-s,3-\Delta_\epsilon)-
B(3-\Delta_\epsilon,2\Delta_\epsilon-4-p+s)\Bigr]
\end{multline}
Then, we are ready to extract the $m\sim 0 $ behavior from \ref{eq:mellin}. The only contribution comes from the residue at $s=0$, so that
\begin{multline}\label{Ipcomplete}
I(p)=\frac{2\pi}{2-\Delta_\epsilon}\Bigl[ B(p+2-\Delta_\epsilon,2\Delta_\epsilon-4-p) +\\-B(p+2-\Delta_\epsilon,3-\Delta_\epsilon)-
B(3-\Delta_\epsilon,2\Delta_\epsilon-4-p)\Bigr],
\end{multline}
which for $p=2$ gives $I(2)\simeq-8.4448$.



\bibliography{bib}{}
\bibliographystyle{unsrt}
\end{document}